\documentclass[fleqn,usenatbib]{mnras}

\usepackage{newtxtext,newtxmath}
\usepackage[T1]{fontenc}

\DeclareRobustCommand{\VAN}[3]{#2}
\let\VANthebibliography\thebibliography
\def\thebibliography{\DeclareRobustCommand{\VAN}[3]{##3}\VANthebibliography}

\usepackage{graphicx}	% Including figure files
\usepackage{amsmath}	% Advanced maths commands

\usepackage[dvipsnames]{xcolor}
\newcommand{\tcw}[1]{\textcolor{WildStrawberry}{#1}} %

\usepackage{cancel}
\usepackage{soul}

\title[Dark matter or MOND]{SPARC galaxies prefer Dark Matter over MOND}

\author[M. Khelashvili et al.]{M. Khelashvili,$^{1,2,3,4}$\thanks{E-mail: khelashvili@fias.uni-frankfurt.de}  
A. Rudakovskyi,$^{3}$ 
and S. Hossenfelder$^{4,5}$ 
\\ 
$^{1}$ Goethe Universit\"{a}t, Max-von-Laue Str. 1, Frankfurt am Main, 60438, Germany \\
$^{2}$ Department of Physics, Princeton University, Princeton, NJ 08544, USA \\
$^{3}$ Bogolyubov Institute for Theoretical Physics of the NAS of Ukraine, Metrolohichna Str. 14-b, Kyiv, 03143, Ukraine \\
$^{4}$ Frankfurt Institute for Advanced Studies, Ruth-Moufang-Str. 1, Frankfurt am Main, 60438, Germany \\
$^{5}$ Munich Center for Mathematical Philosophy, 1 Geschwister-Scholl-Platz, D-80539 Munich, Germany
}

\date{Accepted XXX. Received YYY; in original form ZZZ}

\pubyear{2021}

\begin{document}
\label{firstpage}
\pagerange{\pageref{firstpage}--\pageref{lastpage}}
\maketitle

\begin{abstract}
We currently have two different hypotheses to solve the missing mass problem: dark matter (DM) and modified Newtonian dynamics (MOND). In this work, we use Bayesian inference applied to the \textit{Spitzer Photometry \& Accurate Rotation Curves} (SPARC) galaxies' rotation curves to see which hypothesis fares better. For this, we represent DM by two widely used cusped and cored profiles, Navarro-Frenk-White (NFW) and Burkert. We parameterize MOND by a widely used radial-acceleration relation (RAR). Our results show a preference for the cored DM profile with high Bayes factors in a substantial fraction of galaxies. Interestingly enough, MOND is typically preferred by those galaxies which lack precise rotation curve data. Our study also confirms that the choice of prior has a significant impact on the credible interval of the characteristic MOND acceleration. Overall, our analysis comes out in favor of dark matter.
\end{abstract}


%%%%%%%%%%%%%%%%%%%%%%%%%%%%%%%%%%%%%%%%%%%%%%%%%%

%%%%%%%%%%%%%%%%% BODY OF PAPER %%%%%%%%%%%%%%%%%%

\section{Introduction}

We have long known that the observed velocities of stars and gas in most galaxies are in tension with distributions of the visible matter \citep[see, e.g.,][]{Rubin:78, Bosma:1980}. One way to solve this problem is to add the missing mass in the form of dark matter (DM). The alternative is an ad-hoc hypothesis according to which standard Newtonian dynamics is correct only in a limit of accelerations much greater than some characteristic scale $a_0$, while the motions of objects with smaller accelerations are described by modified Newtonian dynamics (MOND) \cite{Milgrom:83}, in the form of the modified inertia \citep{Milgrom:94, Milgrom:22}, or modified gravity \citep{Bekenstein:1984, Milgrom:10, Milgrom:23}; see also extensive reviews \cite{Famaey:12, Milgrom:14}. Notably, analysis of different galaxies has a common best-fit value of the characteristic accelerations, $ a_0\sim10^{-10}\,\text{m}\,\text{s}^{-2}$. MOND is able to predict the ``flat'' rotational curves without (or with adding a much smaller fraction) missing mass. The MOND hypothesis made it possible to  predict the Tully-Fisher relation between galaxy luminosity and ``flat'' rotational curve velocity \citep[see, e.g.][]{Tully:1977, McGaugh:00}. Also, the value of $a_0$ multiplied by the Hubble time is of the order of the speed of light, which may interpreted as a glimpse of some new general physical theory.

While it is an intriguing idea to modify the laws of classical mechanics or gravity, and it makes for an appealing explanation of the galactic rotational curves, MOND faces significant challenges in other astrophysical areas. Pure MOND has difficulties in explaining gravitational lensing in clusters, e.g., the ``Bullet cluster'' (\cite{Clowe:04}; however see \cite{Angus:06}). Gravitational relativistic theories with a MOND regime also have difficulties fitting the cosmic microwave background \citep[see, e.g.,][]{Dodelson:11}.

In \cite{McGaugh2016}, it was suggested that rotational velocities of the galaxies from the \textit{Spitzer Photometry and Accurate Rational Curve}  (SPARC) catalogue \citep{Lelli2016} follow the tight radial-acceleration relation (RAR). This relation includes only one parameter, characteristic acceleration $a_0\simeq 1.2 \cdot 10^{-10}\,\text{m}\,\text{s}^{-2}$, and may be a manifestation of MOND.
However, \cite{Rodrigues:18} reported very strong (up to $10\sigma$) incompatibility between confidence intervals of characteristic accelerations $a_0$ for individual late-type galaxies from the SPARC catalogue. This led to an extensive discussion of the analysis of galactic rotational curves and ways to improve its reliability.  Independent teams applied different methodologies. Some of them found \citep{Lelli2017, McGaugh2018, Li2018, Desmond:23b}, some did not find \citep{Chang2019, Rodrigues2019, Marra2020, zhou2020absence, Li_Zhao2021} evidence of the presence of universal acceleration scales in {\sc SPARC} galaxies. 

\cite{McGaugh2018} criticized the priors chosen for luminous matter in \cite{Rodrigues:18} and suggested that the statistical distribution found for $a_0$ does not contradict MOND. \cite{Kroupa:18} suggested that the $a_0$ scatter may be caused by the inclination angle of galaxies to line-of-sight and poor quality data. \cite{Cameron:20} suggested that Bayesian techniques should be applied more cautiously, e.g., by using credible intervals for $a_0$ instead of confidence ones (based on $\chi^2$ statistics). On the other hand, \citep{Rodrigues:20, Marra2020} found that the strong tension between $a_0$ arises in the analysis of individual galaxies even if more realistic and physically motivated priors are applied together with credible intervals instead of confidence ones.

In this work, we want to look at this question from a somewhat different angle, by investigating whether DM or RAR without DM is better for describing galaxy rotation curves (RC). For this, we need to take into account that models for DM density profiles usually have more parameters and are more flexible than MOND, which makes it easier to fit data. Therefore, we will follow \cite{Khelashvili:22} and apply Bayesian model comparison for model selection. This approach penalizes models with redundant complexity. We also study the impact of different physically motivated prior choices on the results of out Bayesian inference.

This paper is organised as follows: in Sec.~\ref{sec:model} we describe the dark matter profiles under consideration, radial-acceleration relation and {\sc SPARC} rotational curve model. In Sec.~\ref{sec:analysis} we briefly explain the statistical model and describe our results in Sec.~\ref{sec:results} In Sec.~\ref{sec:discussion} we discuss our results, and we conclude in Sec.~\ref{sec:conclusion}. Throughout this paper we use the Hubble parameter $H_0=73\,\frac{\text{km/s}}{\text{Mpc}}$.

\section{Model}\label{sec:model}

In the {\sc SPARC} database, the total acceleration from the gravitational attraction of baryons is a sum of contributions of the different components (disk, gas, bulge) of a galaxy and parameterized through the ``velocity'' contributions of each component:

\begin{equation}
     a_b = \left(\left|v_\text{gas}\right|v_\text{gas} + \Upsilon_\text{disk} \left|v_\text{disk}\right|v_\text{disk} + \Upsilon_\text{bulge}\left|v_\text{bulge}\right|v_\text{bulge} \right)/r\,
\end{equation}
where $\Upsilon_i$ are luminosity-mass ratios of the components.

In MOND, the physical acceleration is defined as:
\begin{equation}
    a(r) = a_b(r) \nu\left(\frac{a_b}{a_0}\right)\,
\end{equation}
with the asymptotic behavior 
\begin{equation}
    a = 
    \begin{cases}
     a_b, & \text{for} \quad a_b \gg a_0 \\
     \sqrt{a_ba_0}, & \text{for} \quad a_b \ll a_0
    \end{cases}
    \label{mond-empirical-law}
\end{equation}
where $a_b$ is the acceleration from the Newtonian gravitational force created by the baryonic content of the galaxy:
\begin{equation}
    a_b(r) = \frac{G_N M_b(r)}{r^2}\,\tcw{.}
\end{equation}
If MOND is a universal new law of physics, we expect that the characteristic acceleration scale is the same, or almost the same, for different objects.

While the asymptotic behavior of the acceleration in MOND theories is motivated by galactic rotational curves, the exact functional form is not fixed. We here use the RAR as a fiducial interpolation function for MOND \citep[see, e.g.,][]{McGaugh2016}:
\begin{equation}
    a = \frac{a_b}{1 - e^{-\sqrt{\frac{a_b}{a_0}}}}\,\tcw{.}
\end{equation}

We follow the logic of \cite{Khelashvili:22} for dark matter models and consider the Navarro-Frenk-White (NFW) and Burkert density profiles. These profiles are usually described by two parameters: the characteristic density $\rho_s$ and characteristic radius $r_s$. However, for our Bayesian analysis it is more convenient to use a parameterisation in terms of concentration $c_{200}$ and rotation velocity $v_{200}$. Those are related to $\rho_s$ and $r_s$ by:
\begin{equation}
    v_{200} = 10 c_{200}r_s H_0 \;\;\;\; c_{200} = r_{200}/r_s\,,
\end{equation}
where $r_{200}$ is a radius enclosing the average density 200 times greater than the critical density of the Universe.

\section{Analysis}\label{sec:analysis}

Bayes theorem states that the posterior probability density $P\left(\theta\right|D, \mathcal{M})$ of parameters $\theta$ of a model $\mathcal{M}$ is equal to the product of the likelihood $P(D|\theta, \mathcal{M})$  of the data $D$ and priors $\pi(\theta|\mathcal{M})$, divided by the marginalized likelihood (evidence) $P(D|M) = \int P(D|\theta, \mathcal{M})\pi(\theta ) \mathrm{d}\theta$:
$$P\left(\theta\right|D, \mathcal{M}) = \frac{P(D|\theta, \mathcal{M})\pi(\theta|\mathcal{M})}{P(D|\mathcal{M})}.$$
We are interested in the likelihood:
$$P(D|\theta, \mathcal{M})=\prod_\text{i} \frac{1}{\sqrt{2\pi}\sigma_\text{i}}\mathrm{exp}\left(-\frac{\left( v_\text{i,obs}-v_\text{pred}(\theta, \mathcal{M})(r_\text{i}) \right)^2}{2\sigma_\text{i}^2}\right)\,, $$
where $v_\text{i,obs}$ is the observed rotational velocity at radius $r_\text{i}$ (for a galaxy in the SPARC database), $\sigma_\text{i}$ is the error, and $v_\text{pred}(\theta, \mathcal{M})(r)$ is the model $\mathcal{M}$ prediction for rotational velocity.

The Bayes factor $K_{12} = \frac{P(D|\mathcal{M}_1)}{P(D|\mathcal{M}_2)}$, a ratio between the evidences of models $\mathcal{M}_{1}$ and $\mathcal{M}_{2}$, is crucial for model selection. The ratio between posterior probabilities of the models $\mathcal{M}_1$ and $\mathcal{M}_2$ are:
\begin{equation}
    \frac{P(\mathcal{M}_1|d)}{P(\mathcal{M}_2|d)}=\frac{P(d|\mathcal{M}_1)P(\mathcal{M}_1)/P(d)}{P(d|\mathcal{M}_2)P(\mathcal{M}_2)/P(d)}=K_{12}\frac{P(\mathcal{M}_1)}{P(\mathcal{M}_2)}
\end{equation}
If there is no preference either for $\mathcal{M}_1$ or $\mathcal{M}_2$ the preference of model 1 over model 2 reduces to $ \frac{P(\mathcal{M}_1|d)}{P(\mathcal{M}_2|d)}=K_{12}$.

 We use the lognormal distributions for mass-to-light ratios $\Upsilon$ with 0.1 dex dispersion and flat priors for DM parameters similar to \cite{Khelashvili:22}. For the MOND acceleration parameter we will use the priors:
\begin{itemize}
    \item flat prior for $\log_{10}\left(a_0/10^{-13}\text{km}/\text{s}^2\right)$ in range $\left[-3,3\right]$~\footnote{Our flat prior is narrower than in \cite{Marra2020}, which used uniform prior on $\log_{10}a_0$ in diapason $[-20, -5]$.};
    \item gaussian for $a_0/10^{-13} \text{km/s}^2$ with mean $1.2$ and dispersion $0.2$ \citep{Milgrom:14}, which is, however, narrower dispersion found in \cite{Gentile_2011};
    \item fixed value $a_0=1.2\cdot 10^{-13}\text{km}/\text{s}^2$, motivated by the findings of \cite{Marra2020, McGaugh2016}
\end{itemize}
Our choices of priors for the DM and MOND parameters are summarized in Tab.~\ref{tab:models_priors}.

We then apply Bayesian model comparison to the rotational curves from the Spitzer Photometry \& Accurate Rotation Curves (SPARC) database \citep{Lelli2016}. This database includes detailed rotational curves of 175 late-type galaxies based on HI/H$\alpha$ and \textit{Spitzer} infrared observational data. {\sc SPARC} also includes the distributions of luminosity in galaxies, as well as estimates of the contributions of baryonic matter to the gravitational force. In this study, we analyze a subsample of 159 galaxies with quality flag $q = 1, 2$, and $5+$ dots in the rotation curve.
 
To compute the Bayesian evidence, we use the dynamic nested sampling algorithm \citep{Higson:19} implemented in \textsc{dynesty} \textsc{Python} package \citep{dynesty}.
   
\begin{table}
    \centering
    \begin{tabular}{cccccc}
    \hline
    Parameter & Fiducial & Std dev & Units & Range & Prior \\
    \hline
    \multicolumn{6}{c}{Dark matter:} \\
    $v_{200}$ & -- & -- & km / s &  10 -- 500  & uniform \\
    $c_{200}$ & -- & -- & -- & 0 -- 1000 & uniform\\
    \hline
    \multicolumn{6}{c}{MOND (flat):} \\
    $\log_{10} a_{0; -13}$ & -- & -- & 1 & -3 -- 3 & uniform \\
    \hline
    \multicolumn{6}{c}{MOND (gaussian):} \\
    $a_{0; -13}$ & 1.2 & 0.2 & 1 & -- & gaussian \\
    \hline
    \multicolumn{6}{c}{MOND (fixed):} \\
    $a_{0; -13}$ & 1.2 & 0 & 1 & -- & fixed value \\
    \hline 
    \end{tabular}
    \caption{The prior choice for DM and MOND models.}
    \label{tab:models_priors}
\end{table}

\section{Results}
\label{sec:results}

Bayesian model comparison allows us to figure out which model better describes the observed rotational curves.
Our findings are summarized in Tab.~\ref{tab:RAR_vs_DM}, showing comparable numbers of galaxies that favour DM or MOND with RAR interpolation function.  We use the logarithm of the Bayes factor greater or equal to $0.5$, corresponding to the substantial preference, as a criterion for model selection. 
However, when decisive preference threshold of $\Delta \log_{10}Z \ge 2$ is applied, the results strongly favour dark matter. Out of 159 galaxies, 69 show a preference for DM with a Burkert density profile over the RAR with a fixed $a_0$, while only 25 galaxies decisively favour MOND over DM.

Notably, a considerable fraction of galaxies (47 out of 159) prefer a DM Burkert profile over RAR with flat $a_0$ prior with a logarithm of the Bayes factor greater than  If, instead of an individual analysis of galaxies' rotation curve, we would perform a combined fit of RC for the entire sample , the resulting Bayes factor would be a product of the Bayes factors of the individual galaxies. DM would then be strongly favoured over MOND.

\begin{figure*}
      \begin{tabular}{cc}
          \includegraphics[width = 0.5\textwidth]{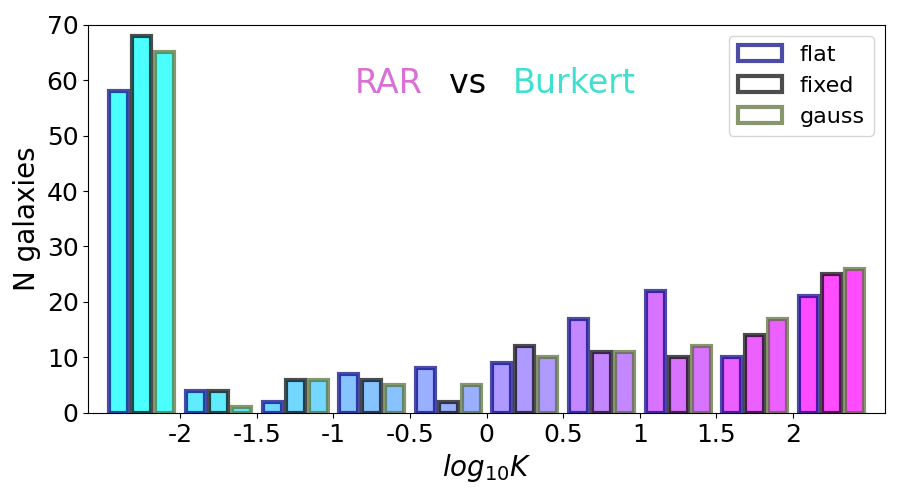}
          &
          \includegraphics[width = 0.5\textwidth]{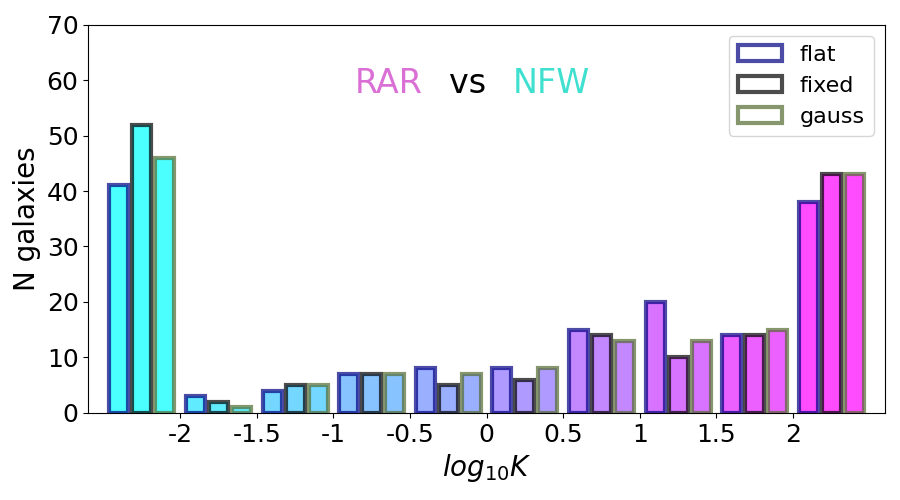}
      \end{tabular}
      \caption{The number of galaxies that prefer (purple) or disfavor (turquoise) MOND RAR compared to DM.}
      \label{fig:rar_vs_dm_3priors}
\end{figure*}

We estimate the prior dependence of our model selection by considering the three different priors on the $a_0$ parameter as listed in Tab.~\ref{tab:models_priors}. The comparison of MOND to DM for different priors is shown in Fig.~\ref{fig:rar_vs_dm_3priors}. The results for all considered priors are close, and the difference between them does not exceed $10$ galaxies in any $\Delta \log Z$ bin. A general number of galaxies preferring DM or MOND are comparable for all priors as well (see Tab.~\ref{tab:RAR_vs_DM}).

We also conduct a similar analysis for the alternative MOND interpolation functions and summarize it in Appendix~\ref{sec:alternative_functions}. We find a similar preference for DM against MOND, which allows us to consider our results as robust.

\begin{table}
    \centering
    \begin{tabular}{l|c|c|c}
     models  & prefer & exclude & indifferent \\
        \hline
        \multicolumn{4}{c}{Flat $\mathrm{log}_{10}a_0$ prior} \\
        \hline
        {\bfseries RAR} vs {\bfseries NFW} &  87 & 56 & 16 \\
        {\bfseries RAR} vs {\bfseries Burkert} & 70 & 72 & 17 \\
\hline
 \multicolumn{4}{c}{Gaussian $a_0$ prior} \\
         \hline
         {\bfseries RAR} vs {\bfseries NFW} & 84 & 60 & 15 \\
        {\bfseries RAR} vs {\bfseries Burkert} & 66 & 78 & 15 \\
         \hline
         
         \multicolumn{4}{c}{Fixed $a_0$} \\
         \hline
         {\bfseries RAR} vs {\bfseries NFW} & 81 & 67 & 11 \\
        {\bfseries RAR} vs {\bfseries Burkert} & 60 & 84 & 15 \\
        
         \hline
    \end{tabular}
    \caption{The number of galaxies that prefer or exclude MOND compared to DM. The total sample contains 159 galaxies.}
    \label{tab:RAR_vs_DM}
\end{table}

\begin{table}
    \centering
    \begin{tabular}{l|c|c|c}
        \hline
        Model & prefer & exclude & indifferent \\
        \hline
         RAR (flat) & 57 & 84 & 18 \\
         RAR (fixed) & 52 & 94 & 13 \\
         RAR (gauss) & 54 & 86 & 19 \\
         \hline
         NFW & 24 & 121 & 14 \\
         Burkert & 53 & 85 & 21 \\
        \hline
    \end{tabular}
    \label{tab:best_model_by_evidences}
    \caption{The number of galaxies that prefer or exclude each model compared to all other considered models. DM models are compared to RAR with flat prior. The threshold for model preference is $\log_{10}Z > 0.5$.}
\end{table}

\subsection{Astrophysical properties of galaxies preferring MOND}

\begin{figure*}
    \includegraphics[width = 0.95\textwidth]{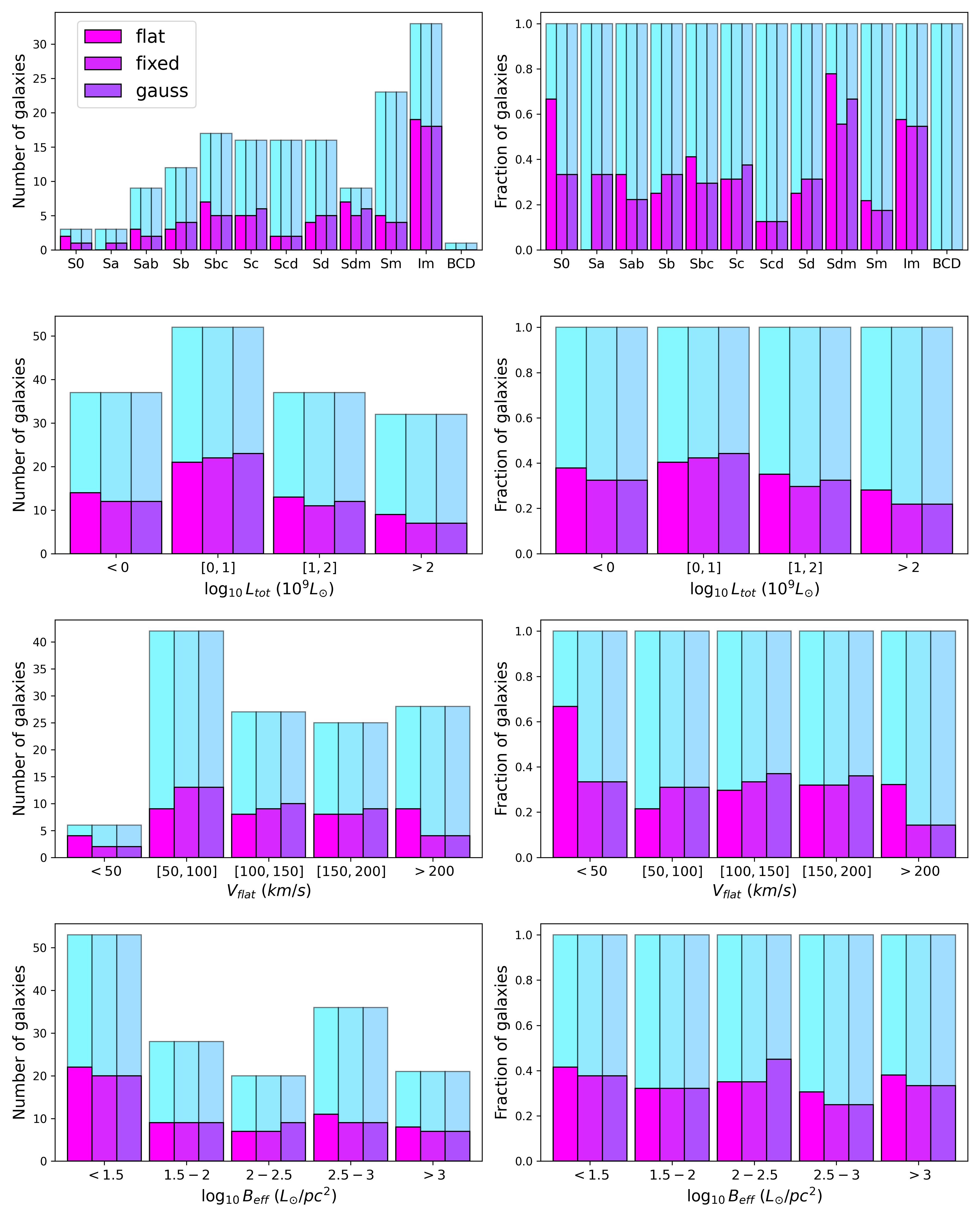}
    \label{fig:correlations}
    \caption{Correlation between preference of MOND (RAR) with flat $a_0$ prior and astrophysical parameters of galaxies.}
\end{figure*}

Another question of interest is exploring the correlation between the astrophysical properties of galaxies and the preferred rotational curve model.
We consider galaxies' morphological type, flat velocity (that correlates with the galaxy mass), total luminosity, and surface brightness as the astrophysical characteristics of the galaxies. These parameters are given in the SPARC database and are model-independent. The fractions of galaxies that substantially prefer MOND over DM in different spans of astrophysical parameters are shown in Fig.~\ref{fig:correlations}. Overall, the fraction of galaxies preferring MOND is close for all the ranges of considered parameters, with the only exemption of galaxies of the diffuse and irregular morphological types (Sdm, Irr) that prefer MOND in more than half of cases. 

\subsection{Rotational curve features of galaxies preferring MOND}

The {\sc SPARC} database contains galaxy rotation curve data combined from different sources, and the ``quality'' of the data significantly varies from galaxy to galaxy \cite[see more in][]{Lelli2016}. 
The rotation curve features, such as inclination angle, number of data points and relative values of observational errors may impact the model selection. For example, underestimated errors may lead to the preference of more complicated models, that have enough degrees of freedom to adjust all RC features. At the same time, models with lower complexity may benefit from the data with less constraining power.

\begin{figure*}
    \includegraphics[width = 0.9\textwidth]{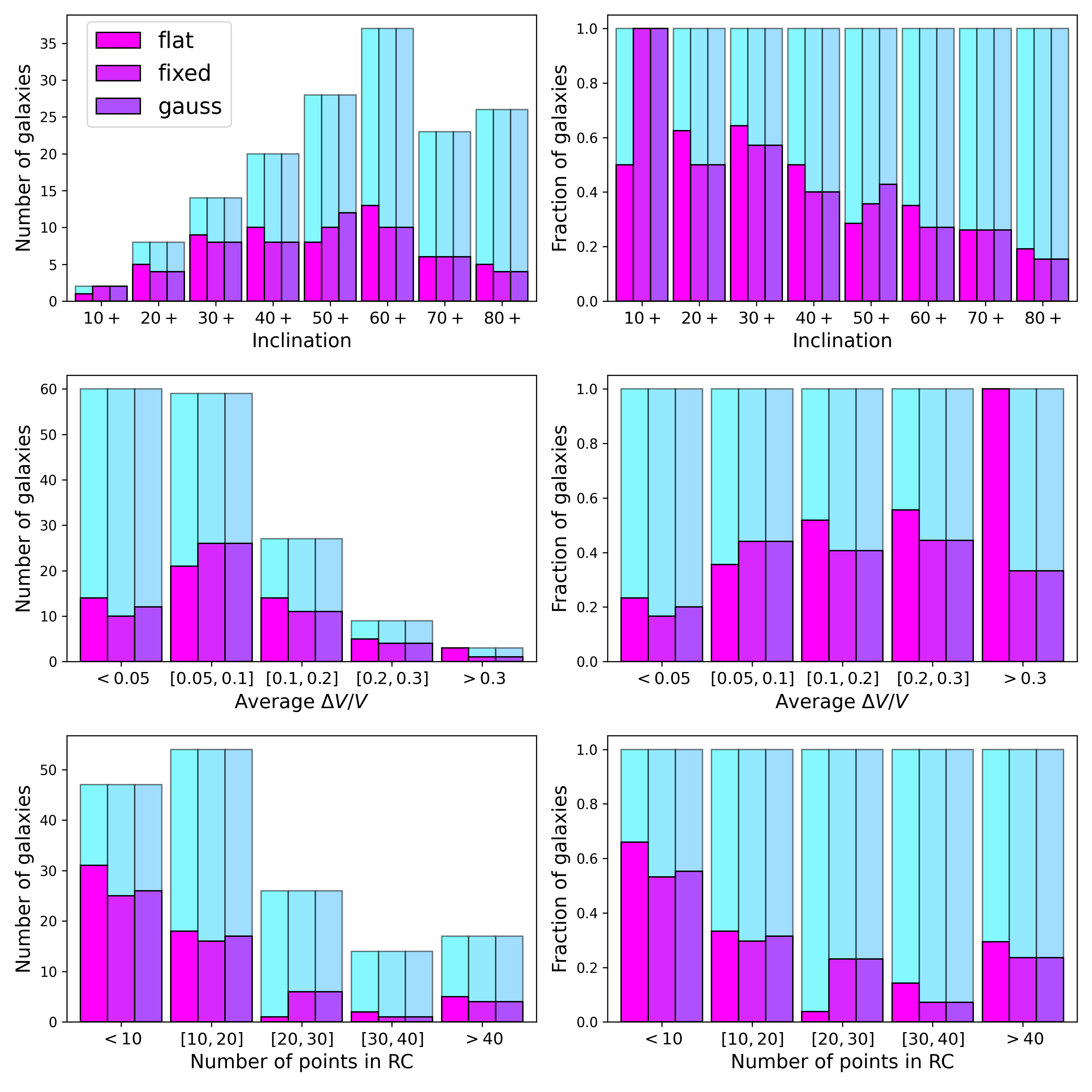}
    \label{fig:quality_correlations}
    \caption{Correlation between preference of RAR with flat $a_0$ prior and parameters that may impact data quality.}
\end{figure*}

First, we find that the fraction of galaxies that prefer MOND is higher among galaxies with inclination angle $i<30^{\circ}$, see Fig.~\ref{fig:quality_correlations}. %for more details.
For such nearly face-on galaxies, the degeneracy between the astrophysical and halo parameters is especially relevant. 
At the same time, the test of inference on the subset of galaxies with $i>30^{\circ}$ and best-quality data reveals approximately the same fraction of galaxies that prefer or reject MOND as in an overall set of galaxies.

\begin{figure}
    \centering
    \includegraphics[width = 0.5\textwidth]{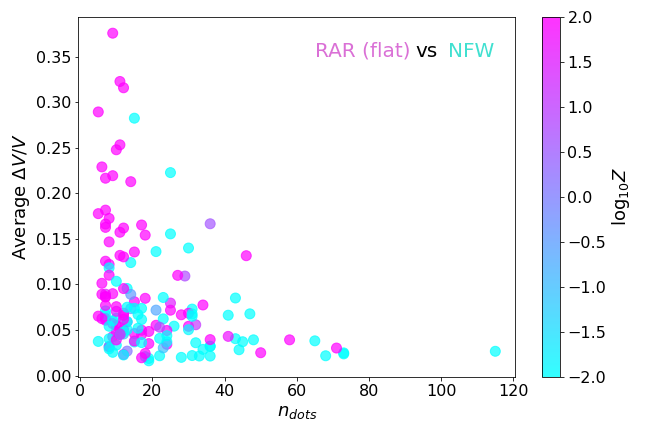}
    \caption{
    The distribution of galaxies preferring RAR (flat prior) or NFW on the diagram of the number of dots in the RC vs. the average error of the observed velocity. Similar correlations are present for all DM models vs. RAR with any priors; however, the shown pair of models has the most prominent separation.}
    \label{fig:ndot_average_error_correlation}
\end{figure}

Another interesting observation is that there are almost no galaxies 
in our sample which simultaneously have a large number of points in RC and large 
average errors in observed velocity. We see a tendency that galaxies with a small number of data points ($<20$) are better fitted with MOND and that DM is a better fit for galaxies with a smaller ($<5\%$) average error and larger ($>20$) number of points in the rotation curves,
see Fig.~\ref{fig:ndot_average_error_correlation}.
Only a small number of galaxies ($6$ galaxies for flat $a_0$ prior RAR vs. NFW) prefer MOND over DM among the galaxies that have more than 20 data points in the RC and the average errors smaller than $5\%$. That could be an indication that the preference for MOND is due to the penalty for the number of free parameters in the Bayesian inference from the galaxies whose RC data is not sufficient to distinguish the models in terms of likelihood. 

\subsection{MOND characteristic acceleration}

\begin{figure*}
    \centering
    \includegraphics[width = \textwidth]{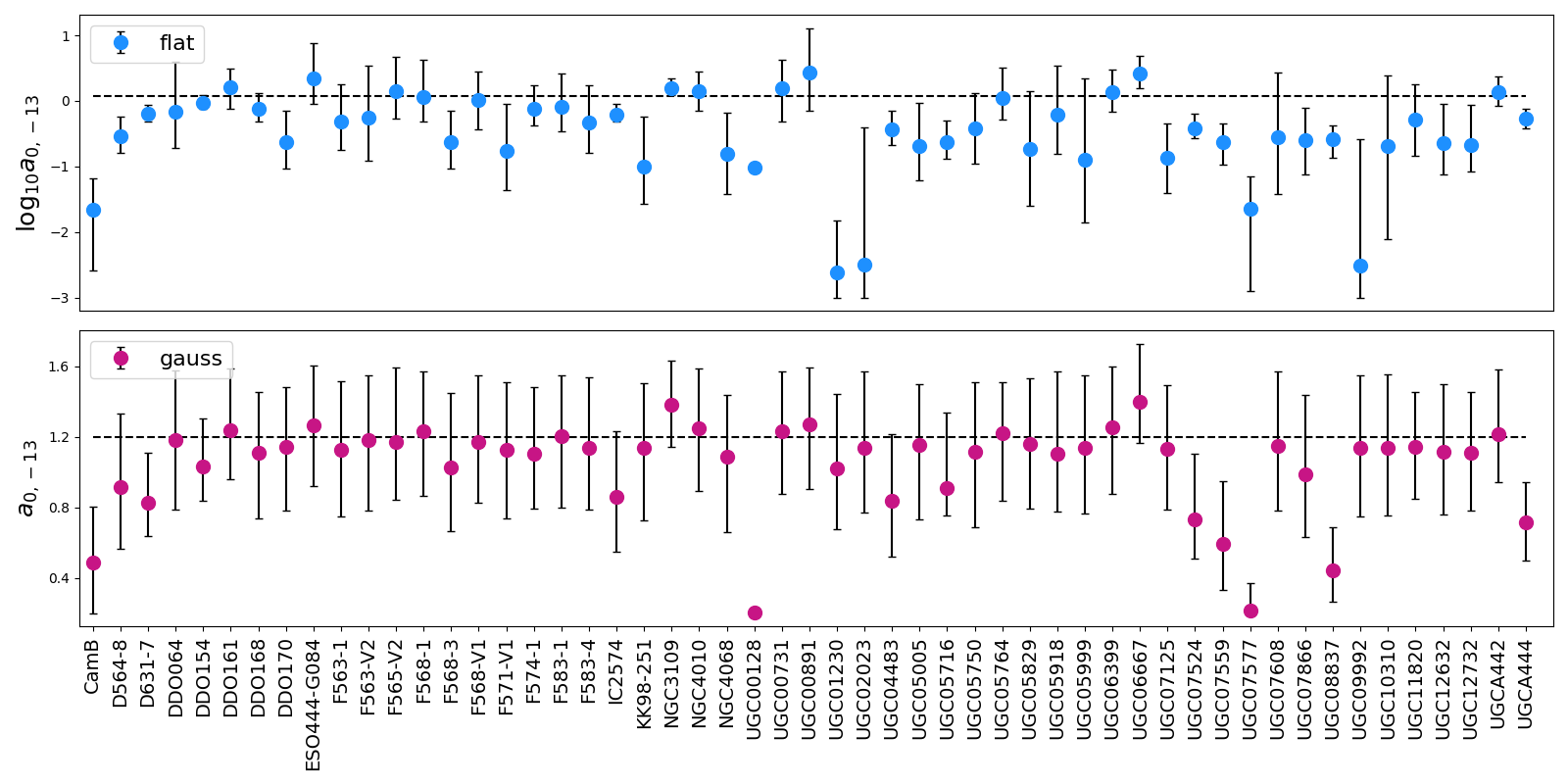}
    \caption{Credible $95\%$ region for the MOND characteristic acceleration parameter $a_0$ for the sub-sample of LSB galaxies from SPARC. The upper plot shows a credible region for MOND analysis with a flat $a_0$ prior and the lower one for the Gaussian $a_0$ prior.}
    \label{fig:rar-a0-cred}
\end{figure*}

Let us now look at the consistency of the preferred values of MOND characteristic acceleration among the studied galaxies. 
The results for the $a_0$ mean values and $2\sigma$ credible regions for low surface brightness (LSB) galaxies are given in Fig.~\ref{fig:rar-a0-cred}. While for most of the galaxies, Milgrom's $a_0 = \ 1.2 \cdot 10^{13} \text{km}/\text{s}^2$ lies within the $2\sigma$ region, there are 24 of 53 (flat priors) and 8 of 53 (Gaussian prior) for which this is not the case. In addition, $2\sigma$ regions of some galaxies disagree with each other. However, the degree to which the credible intervals disagree with standard $a_0$ value is noticeably different for the model with flat and Gaussian prior on $a_0$. In the latter case, these are, as one expects, more uniform and, in most cases, contain the standard value. Moreover, the fraction of galaxies preferring MOND with Gaussian prior and even fixed value of $a_0$ is only slightly smaller than for the flat prior case. 

\subsection{Bayesian information criterion model selection}

As an independent estimate, we compared the models based on the Bayesian information criterion (BIC):
\begin{equation}
    \text{BIC} = -2\log\mathcal{L}(D|\theta_{\text{ML}}, M) + p\log n\,,
    \label{bic}
\end{equation}

We consider the ``extended'' BIC, which takes into account the penalty from priors of baryonic parameters (that directly follow from SPARC estimations and stars population model):
\begin{equation}
    \text{BIC}^\text{ext} =  2\log \mathcal{L}(D|\theta_{MAP}) - 2\log\pi(\theta_{astro, MAP}) + p\log n_*\,,
\end{equation}
here $n_* = n + p_\text{astro}$ is total number of ``observational points'' including priors for the astrophysical parameters. The details on the motivation of such extension and the derivation of this expression are given in the App.~\ref{sec:bic-app}.

\begin{figure*}
    \centering
    \begin{tabular}{cc}
        \includegraphics[width = 0.5\textwidth]{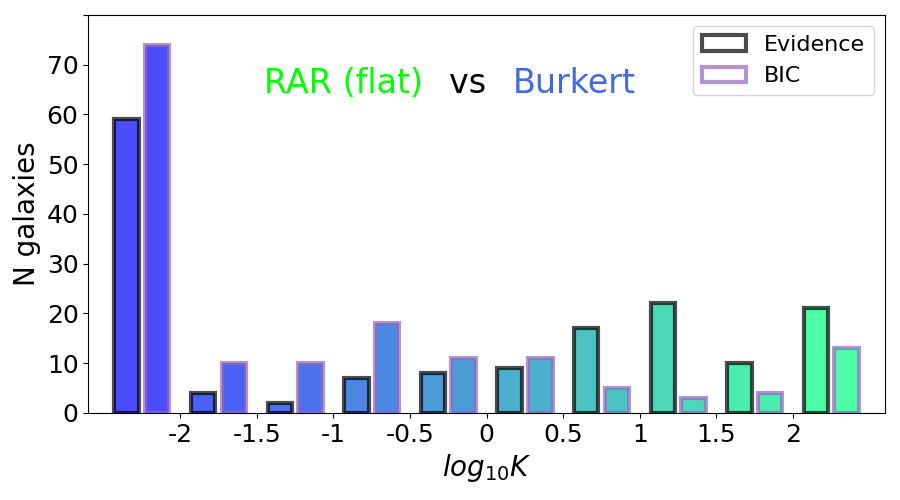} &
        \includegraphics[width = 0.5\textwidth]{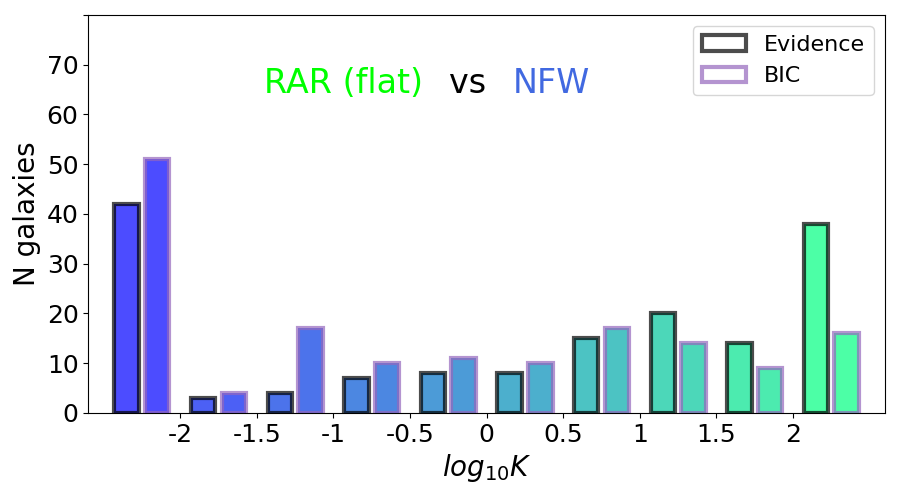} \\
        \includegraphics[width = 0.5\textwidth]{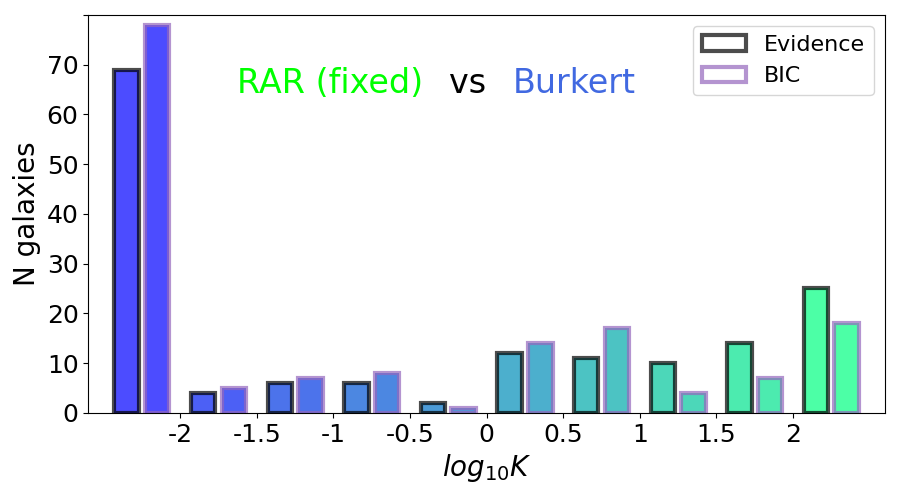} &
        \includegraphics[width = 0.5\textwidth]{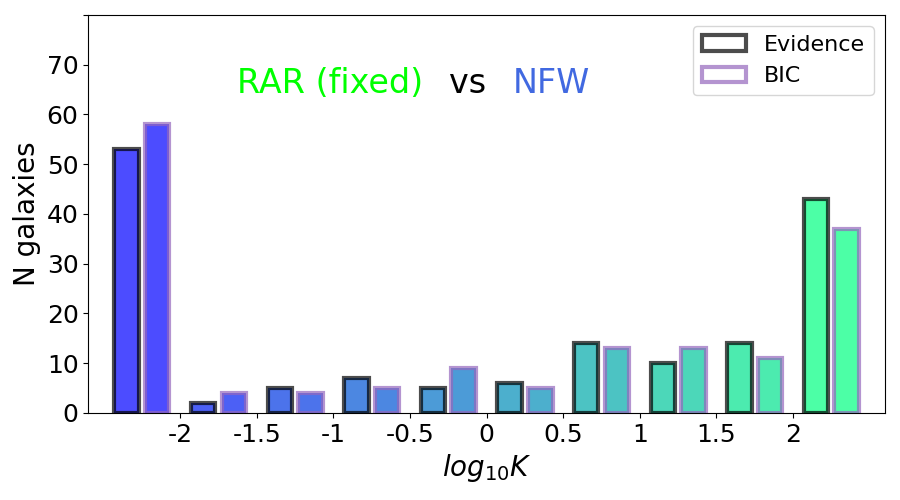} 
    \end{tabular}
    \caption{Comparison of model preferences based on evidence inferred by nested sampling algorithm and evidence estimated by BICs.}
    \label{fig:bics-evidences}
\end{figure*}

The Bayes factors estimated by extended BIC are compared with Bayes factors calculated by  nested sampling in Fig.~\ref{fig:bics-evidences}. The results based on the two methods are close, and we can conclude that the impact of prior volume and choice of methodology play a minor role in the performed analysis. Other common suggestions are addressed in the Discussion.

\section{Discussion}\label{sec:discussion}
We now want to compare our results with the findings of previous analyses. Several groups reported that DM may provide a better fit for SPARC galaxies than MOND with the RAR interpolation function. \cite{wang2020comparison} suggested that the combined fit of the SPARC galaxies with DM Burkert profile gives better reduced $\chi^2$ than with the RAR. The distributions of additional velocities \tcw(the difference between the observational rotation curve and the expected one from the baryonic matter) seem to be in a better agreement with predictions of Burkert and DC14 DM profiles than RAR ones \cite{Hernandez2022}. Also, \cite{Mercado:23} reported that  $\sim15$ \% galaxies in the SPARC sample exhibit acceleration behaviour which is difficult to explain in the MOND paradigm, but could be explained in the dark matter-dominated system with a cored density profile.

 The scatter of the $a_0$ credible intervals that we found with a flat $a_0$ prior is in good agreement with \cite{Marra2020}. The strong degeneracy between parameters, which may arise with wide flat priors, could lead to misleading results \citep[see, e.g.][]{Li:21a}. To minimize this effect, we exploit the Gaussian prior for $a_0$. We find that the discrepancy between the $a_0$ credible intervals is much smaller in this case compared to the ones obtained within the flat $a_0$ priors. Qualitative similar result was reported in the \cite{Chang2019}, which used the Gaussian prior on $\log_{10}a_0$ with $\mu = -12.9$ and much wider dispersion $\sigma = 2$ than ours. 

If the RAR is a consequence of MOND, it should be manifested on the wide range of masses of cosmological objects. The ultra-faint and classical dwarf spheroidal galaxies seem to be following the RAR \citep[see, e.g.][]{Lelli2017, Eftekhari:22} as well as disk galaxies. However, the acceleration constants found within the analysis of different galaxy clusters observed by the  X-ray telescopes deviate by more than one magnitude from RAR expectations \citep{Chan:20, Pradyumna:21}.
Some galaxy groups also do not follow the RAR \citep{Gopika:21}.

It has furthermore been found that RAR-like relations may naturally arise in the standard Cold Dark Matter model (\cite{Keller:17, Ludlow:17, Dutton:19, Paranjape:21, Mayer:23, Mercado:23}; however see \cite{Milgrom:16, Li:21b}). This might be a hint as for why MOND works better than DM for a significant fraction of galaxies in the sample. 

In our work, we focus on the most popular and simplest version of MOND. However, it is well-known that this RAR interpolation function is strictly speaking only correct for spherically symmetric systems.
Our approach also does not take into account the possible external field effects. These effects, caused by the gravitational field of the environment, may play a significant role; e.g. \cite{Chae:20, Chae:21, Chae:22} reported evidence of external field effect in SPARC galaxies. \cite{Chae:22b} suggested that the modified gravity works better than modified inertia.

The {\sc SPARC} database provides the best rotational curves and photometry among existing ones. However, there is a concern that observed rotational curves do not accurately represent the circular velocities of test particles. The real star and gas motions in galaxies may be different from circular ones and baryons may be not in equilibrium in galaxies (S. McGaugh, private communications). Also, \cite{Roper:23} found that due to these effects, dwarf galaxies' density profiles may seem more flattened than they really are. We hope to address the detailed analysis of the impact of these effects on the  DM/MOND model selection in future work.

Our work also faces the limitations of applied statistical techniques. 
Though Bayesian inference is the most reliable method for model selection and is recommended to apply whenever possible \citep[see, e.g.,][]{vonToussaint2011}, this method is sensitive to the prior choice. For parameter estimation, prior should not introduce any bias on the parameters' values and should be maximally wide (uninformative), if it is not strictly defined from the previous observations. This leads to freedom in the prior choice. This freedom could impact significantly the preferred physical model (which is not the same as a statistical model). 
This can be shown as follows. If $\mathcal{L}(D|\theta_{\text{ML}}, M)$ is a likelihood estimator, $p$ is the number of parameters, $(\Delta\theta_{\text{prior}})^p$ and $(\Delta\theta_{\text{like}})^p$ are the prior volume and the region of the parameter where the likelihood is substantially larger than zero, the evidence may be roughly estimated as $P(D|M) \sim \mathcal{L}(D|\theta_{\text{ML}}, M) \left(\frac{\Delta\,\theta_{\text{like}}}{\Delta\,\theta_{\text{prior}}}\right)^p \,
$. Thereby, for example, tripling the suitable prior range for one parameter introduces the factor of $\sim 10^{-0.5}$ to the Bayes factor, which corresponds to substantial model preference. Though several approaches, like  Jeffrey's \citep{Jeffreys:1946}  or reference priors \citep{Berger:09}, were proposed to overcome this problem, they also have some shortcomings, e.g., they are often improper (cannot be normalized). Jeffrey's prior also works properly only for one-dimensional cases. To check our results, we use different priors as well as the Bayesian information criterion. BIC is an estimator of the Bayesian evidence and doesn't depend on the prior choice by definition. We address the search for the best priors for the galaxy rotational curve analysis for future work.

\section{Conclusions}\label{sec:conclusion}

We here compared two competing hypotheses that have emerged as responses to the missing mass problem within galaxies---dark matter and modified Newtonian dynamics---by using a Bayesian framework to analyse how well each of them explains the rotational curves of galaxies.

Our comparison of the Bayesian evidence for DM and MOND has revealed that the strength of preference in individual galaxies is distributed differently for each model. DM is more often a decisively better fit for individual galaxies, while MOND-fits are more uniformly good or bad. As a result, while the total numbers of galaxies which are better fit by DM (Burkert) or MOND are comparable, there is a considerable fraction of galaxies for which the DM fit is much better than (with a high Bayes factor $\log_{10} K > 4$). As a consequence, DM comes out ahead overall.  

We further found that galaxies with relatively large measurement errors and a small number of data points tend to be better fitted with MOND. There are only a few galaxies with a small average error 
(less than $5\%$) and a sufficiently large number of points in the rotation curves (more than $20$) that are better fitted with MOND than with DM. This suggests that a preference for MOND may come from galaxies whose rotation curve data is insufficient to distinguish the models, in which case MOND is preferred by the Bayesian analysis because it has fewer free parameters.
 
We double-checked our results by studying several RAR characteristic acceleration priors and exploiting the alternative  Bayesian information criterion (BICs). We confirmed that our results are robust and do not show any significant dependence on the prior choice or methodology.  

Our analysis showed that credible intervals of RAR characteristic acceleration are strongly dependent on the prior choice. 
The tension between the credible intervals on $a_0$ preferred by different galaxies is almost eliminated by imposing a more suitable (gaussian) prior on $a_0$. Moreover, the choice of stricter prior or even fixing $a_0$ does not alternate much the number of galaxies preferring MOND. At the same time, the number of galaxies preferring DM (Burkert) is higher than for RAR with a fixed $a_0$ value. 

In summary, our analysis favors Dark Matter. However, the significant degeneracy between astrophysical and cosmological parameters, coupled with concerns regarding whether the observed rotational curves faithfully represent the circular motions of test particles in the gravitational field of galaxies, possible external field effects, and limitations in existing statistical approaches, preclude us from conclusively rejecting MOND.

\section*{Acknowledgements}

We are grateful to Stacy McGaugh, Tobias Mistele, Mariangela Lisanti, Luciano Rezzolla and Yuri Shtanov for the fruitful discussions. The work of AR is supported by the National Research Foundation of Ukraine under Project No.~2020.02/0073 and by the Simons Foundation. MK acknowledges support from  GRADE starting scholarships for international doctoral candidates at Goethe University and from the Department of Energy~(DOE) under Award Number DE-SC0007968. Also, MK is grateful to Princeton University and Frankfurt Institute for Advanced Studies for their hospitality. The calculations presented here were performed on the BITP computer cluster.

\section*{Data availability}

The SPARC data we analyze in this paper are publicly available. We will share the code developed and used in this work upon a reasonable request to the corresponding author.



\bibliographystyle{mnras}
\bibliography{main}

%\begin{thebibliography}{99}
%\end{thebibliography}

%%%%%%%%%%%%%%%%%%%%%%%%%%%%%%%%%%%%%%%%%%%%%%%%%%

%%%%%%%%%%%%%%%%% APPENDICES %%%%%%%%%%%%%%%%%%%%%

\appendix

\section{Simple and standard MOND interpolation functions vs dark matter}\label{sec:alternative_functions}
In complement to RAR, two other MOND interpolation functions are studied, the ``simple'' one \citep{Famaey_2005}: 
\begin{equation}
    a = a_\text{bar}\left( \frac{1}{2} + \sqrt{\frac{a_0}{a_\text{bar}} +\frac{1}{4}} \right) \,,
\end{equation}
and ``standard'' one \citep{McGaugh:08}:
\begin{equation}
    a = a_\text{bar} \sqrt{\frac{1}{2} + \sqrt{\frac{a_0^2}{a_\text{bar}^2} + \frac{1}{4}} }\,.
\end{equation}

The results for the model comparison for these interpolation functions are described in Tab.~\ref{tab:Simple_vs_DM} and Tab.~\ref{tab:Standard_vs_DM} and are generally close to the RAR.
Interestingly, with the flat $a_0$ prior MOND with RAR and simple interpolation functions fit the data %comparably well and 
better than standard, and in the case of fixed $a_0$, simple turned out to be the best interpolation function. 
\begin{table}
    \centering
    \begin{tabular}{l|c|c|c}
     models  & prefer & exclude & indifferent \\
        \hline
        \multicolumn{4}{c}{Flat $\mathrm{log}_{10}a_0$ prior} \\
        \hline
        {\bfseries Simple} vs {\bfseries NFW} &  86 & 56 & 17 \\
        {\bfseries Simple} vs {\bfseries Burkert} & 70 & 73 & 16 \\
        \hline     
         \multicolumn{4}{c}{Fixed $a_0$} \\
         \hline
         {\bfseries Simple} vs {\bfseries NFW} & 88 & 57 & 14 \\
        {\bfseries Simple} vs {\bfseries Burkert} & 71 & 73 & 15 \\
         \hline
    \end{tabular}
    \caption{The number of galaxies that prefer or exclude MOND with the simple interpolation function compared to DM. The total sample contains 159 galaxies.}
    \label{tab:Simple_vs_DM}
\end{table}

\begin{table}
    \centering
    \begin{tabular}{l|c|c|c}
     models  & prefer & exclude & indifferent \\
        \hline
        \multicolumn{4}{c}{Flat $\mathrm{log}_{10}a_0$ prior} \\
        \hline
        {\bfseries Standard} vs {\bfseries NFW} &  84 & 63 & 12 \\
        {\bfseries Standard} vs {\bfseries Burkert} & 69 & 78 & 12 \\
        \hline   
         \multicolumn{4}{c}{Fixed $a_0$} \\
         \hline
         {\bfseries Standard} vs {\bfseries NFW} & 78 & 71 & 10 \\
        {\bfseries Standard} vs {\bfseries Burkert} & 59 & 84 & 16 \\
         \hline
    \end{tabular}
    \caption{The number of galaxies that prefer or exclude MOND with standard interpolation function function compared to DM. The total sample contains 159 galaxies.}
    \label{tab:Standard_vs_DM}
\end{table}

\section{BIC as a marginal likelihood estimator}
\label{sec:bic-app}

Here, we discuss in detail how BIC should be extended to account for the reasonableness of the astrophysical parameters in a way that its interpretations as an evidence estimator remain valid.

To account for astrophysical parameters, we extend the likelihood so that it contains astrophysical parameters priors: 
\begin{equation}
    \mathcal{L}_\text{ext}(D|\theta, M) = \mathcal{L}(D|\theta, M)\times \pi(\theta_{\text{astro}})\,.
\end{equation}
The extended BIC bases on this likelihood redefinition:
\begin{equation}
    \text{BIC}_\text{ext} = -2\log\mathcal{L}_\text{ext}(D|\theta, M) + p\log n_*\,,
    \label{new-bic}
\end{equation}
now the effective number of observational points increases by the number of astrophysical parameters $n_* = n + p_\text{astro}$.
And since all the DM and MOND priors are flat (except the gaussian $a_0$ prior, which is not considered here):
\begin{equation}
    \mathcal{L}_\text{ext}(D|\theta_{\text{ML}}, M) = \mathcal{L}(D|\theta_{\text{MAP}}, M) \times \pi(\theta_{\text{astro, MAP}})\,,
\end{equation}
the maximal likelihood (ML) of the extended likelihood corresponds to the maximum a posteriori (MAP) for the conventionally defined quantities, i.e. maximizes the likelihood and priors product.

It is worth noting that such an extension of BIC is legitimate, as the same prior terms as in Eq.~\eqref{new-bic} appear in the derivation of BIC as a marginal likelihood estimator and are usually neglected because of the smallness.

With the assumption that the evidence integral is concentrated around ML and using the Laplace approximation for this integral, one receives the following expression for BIC as an evidence estimator \citep{konishi2008}:
\begin{multline}
    \text{BIC} = -2\log Z \approx -2\log \mathcal{L}(D|\theta_{ML}) - 2\log\pi(\theta_{ML}) + \\ p\log n - p\log(2\pi) + \log |J(\theta)|\,,
    \label{konishi-full-bic}
\end{multline}
where
\begin{equation}
J(\theta) = \frac{1}{n}\frac{\partial^2\log\mathcal{L}}{\partial \theta\partial\theta^T}\,.   
\end{equation}

In general case, the under integral function maximizes at $\theta_\text{MAP}$ that may not coincide with ML.
By replacing $\theta_\text{ML} \to \theta_\text{MAP}$ and combining  first two terms $ \mathcal{L}\times \pi_\text{astro} \to \mathcal{L}_\text{ext}$ from Eq.~\ref{konishi-full-bic} we would have:
\begin{multline}
    \text{BIC} = -2\log Z \approx -2\log \mathcal{L}(D|\theta_{MAP}) - 2\log\pi(\theta_{astro, MAP})  \\ - 2\log\pi(\theta_{model, MAP}) + p\log n_* - p\log(2\pi) + \log |J(\theta)|\,,
    \label{bic-inermediate}
\end{multline}

We neglect the last term in Eq.~\ref{bic-inermediate} as in the original BIC derivation and remove the model (MOND or DM) priors as this was a final goal of considering BIC (this is equivalent to the assumption of well-chosen model's priors diapasons $\Delta\theta_\text{prior} \sim \Delta\theta_\text{like}$): 
\begin{equation}
    \text{BIC}_\text{ext} = 2\log \mathcal{L}(D|\theta_{MAP}) - 2\log\pi(\theta_{astro, MAP}) + p\log n\,.
\end{equation}
The last expression we use as a definition of extended BIC. 

Our extension of BIC is valid when the peak of the posterior probability is sharp enough, and the Laplace approximation for integral can be applied similarly to the original derivation of BIC as an evidence estimator.

\begin{figure}
    \centering
    \includegraphics[width = 0.5\textwidth]{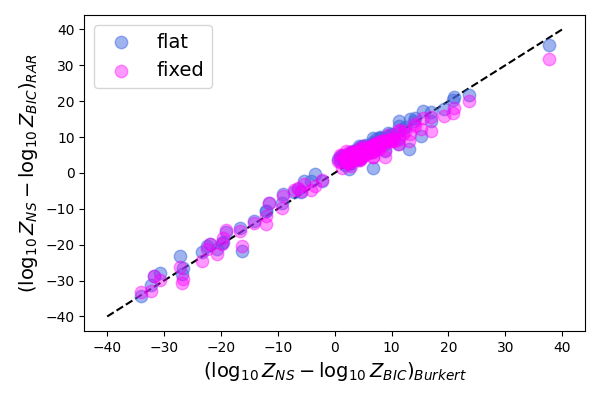}
    \caption{The difference between the evidence calculated by the nested sampling algorithm and estimated by BIC for RAR versus dark matter Burkert profile. }
    \label{fig:bic-evidence-rar-dm}
\end{figure}

We should note that for studied galaxies sample the BICs approximate evidence very roughly, and the typical difference between the evidence calculated by nested sampling algorithm with the evidence estimated by BICs (as in the Eq.~\ref{bic}) is larger than the usual strong preference threshold.
Meanwhile, the deviations from the actual evidence correlate between all the models (see Fig.~\ref{fig:bic-evidence-rar-dm}). So, this approximation can be used for the model comparison with caution.

%%%%%%%%%%%%%%%%%%%%%%%%%%%%%%%%%%%%%%%%%%%%%%%%%%

% Don't change these lines
\bsp	% typesetting comment
\label{lastpage}
\end{document}